\documentclass[12pt]{article}

\begin{document}

\title{Cosmological bulk viscosity, the Burnett regime, and the BGK equation}
\author{A. Sandoval-Villalbazo$^a$ and L.S. Garc\'{\i}a-Col\'{\i}n$^{b,\,c}$ \\
$^a$ Departamento de Ciencias, Universidad Iberoamericana \\
Lomas de Santa Fe 01210 M\'{e}xico D.F., M\'{e}xico \\
E-Mail: alfredo.sandoval@uia.mx \\
$^b$ Departamento de F\'{\i}sica, Universidad Aut\'{o}noma Metropolitana \\
M\'{e}xico D.F., 09340 M\'{e}xico \\
$^c$ El Colegio Nacional, Centro Hist\'{o}rico 06020 \\
M\'{e}xico D.F., M\'{e}xico \\
E-Mail: lgcs@xanum.uam.mx}
\maketitle

\begin{abstract}
Einstein\'{}s field equations in FRW space-times are coupled to the BGK
equation in order to derive the stress energy tensor including dissipative
effects up to second order in the thermodynamical forces. The space-time is
assumed  to be matter-dominated, but in a low density regime for which a
second order (Burnett) coefficient becomes relevant. Cosmological
implications of the solutions, as well as the physical meaning of transport coefficients in an isotropic homogeneous universe are discussed.
\end{abstract}

\section{\textbf{Introduction}}

The physics of low density systems is relevant in the description of the late universe.
Since a hydrodynamical, Navier-Stokes, description of a low density fluid
breaks down for small Knudsen numbers, it is desirable to go beyond this
regime and analyze the fluid with the tools that statistical physics provide
as the particle collision frequency decreases. One simple tool useful for
this purpose is the BGK equation, which simplifies the formalism of the full
Boltzmann equation and provides a good grasp about how the non-equilibrium
system evolves. In particular, dissipative effects, and generalized
transport coefficients are derivable from the BGK formalism. These
coefficients can be incorporated to the stress-energy tensor that is, in
turn, coupled to Einstein's field equations.

The relevance of cosmological transport coefficients has already been widely
recognized \cite{Weinberg}, nevertheless, until now, the second order
Burnett equations \cite{GC} \cite{Cow} have never been applied to
isotropic-homogeneous space-times, although the issue has been addressed in
the context of special relativity \cite{Anderson1} \cite{Anderson2}. One
possible motivation for the study of the Burnett-Einstein solutions is the
well-known relation between wave scattering processes, density fluctuations
and dissipative effects \cite{Berne} \cite{Mountain}. Dynamic structure factors, possibly
relevant for describing CMB anisotropies or distortions, can be derived form
the hydrodynamical Burnett approach, as it has already been done for
collisionless plasmas \cite{Yip}-\cite{Yo2}. In this work we analyze the solutions to the BGK equation taking a FRW metric as power series expansions in the relaxation time $ \tau $. To second order, the solution is used to obtain the constitutive equations for the stress tensor thus providing explicit expressions for the transport coefficients. The explicit use of these results in specific situations will be deferred for future applications. Here we will be concerned with their physical and general features. Special emphasis will be placed in the significance of a bulk viscosity in homogeneous and isotropic systems. Unfortunate use of language has given rise to an identification of a geometrical aspect of the universe with a conventional dissipative effect. This gives rise to ambiguities that must be clarified. To accomplish this task, the paper is divided as follows: section 2
reviews the BGK equation \cite{Gross} in FRW space-times and outlines the basic strategy
of its solution up to second order by means of a Chapman-Enskog type
expansion (Burnett regime). Section three, is dedicated to the analysis of
the first and second order distribution functions, the establishment of
expressions for the transport coefficients and the construction of the
Einstein's field equation with a FRW flat metric. Final remarks about the
implications of the Burnett regime in cosmology are included in section four.

\section{BGK equation in FRW space-times}

We start this section writing the standard form of the BGK equation in its
relativistic version for a single component system:
\begin{equation}
\frac{Df}{Dt}=-\frac{f-f^{(0)}}\tau  \label{BGK1}
\end{equation}
where $f=f(x^\mu ,v^\mu )$ is the non-equilibrium distribution
function, $ f^{(0)}$ is the equilibrium function, $\tau $ is the
collision time (mean free time) and $\frac D{Dt}$ is the absolute
derivative with respect to time $t$. The LHS of equation
(\ref{BGK1}) can be written as:
\begin{equation}
\frac{Df}{Dt}=\frac{\partial f}{\partial x^\mu }\,v^\mu
+\frac{\partial f}{
\partial v^\mu }v^\beta \,(\frac{\partial \,v^\mu }{\partial x^\beta }
+v^\alpha \Gamma _{\alpha \beta }^\mu )  \label{TD1}
\end{equation}
Assuming a flat, homogeneous and isotropic space-time, Eq. (\ref{TD1})
reduces to
\begin{equation}
\frac{Df}{Dt}=\frac{\partial f}{\partial x^4}\,v^4+\frac{\partial
f}{
\partial v^4}\,v^\beta \,v^\alpha \,\Gamma _{\alpha \beta }^4  \label{TD2}
\end{equation}
Here we wish to make sure that the ensuing consequences remain clear. Eq. (\ref{TD2}) has no explicit contributions arising from spatial gradients due to the assumption mentioned above. The second term in the R.H.S. containing the non-vanishing Christoffel symbols has an intrinsic geometrical origin so that all effects arising from it must be associated precisely to such structure.
Now, denoting $a(ct)$ the scale factor, $c$ the speed of light, $E$ the
total mechanical energy of a single particle of rest mass $m_o$ , $\gamma $
the usual relativistic factor and $g_{\mu \nu }$ the metric tensor, and taking
into account the expressions:
\begin{equation}
g_{\mu \nu }=\left[
\begin{array}{llll}
a(ct) & 0 & 0 & 0 \\
0 & a(ct) & 0 & 0 \\
0 & 0 & a(ct) & 0 \\
0 & 0 & 0 & -1
\end{array}
\right]  \label{m1}
\end{equation}
\begin{equation}
x^\mu =\left[
\begin{array}{l}
x^1 \\
x^2 \\
x^3 \\
ct
\end{array}
\right]  \label{cvp1}
\end{equation}
\begin{equation}
v^\mu =\left[
\begin{array}{l}
u^1\gamma \\
u^2\gamma \\
u^3\gamma \\
\frac E{m_o\,c}
\end{array}
\right]  \label{cvv1}
\end{equation}
we can get a simpler form of Eq. (\ref{TD2}), namely,
\begin{equation}
\frac{Df}{Dt}=\frac E{m_oc^2}\frac{\partial f}{\partial
t}\,+m_o\frac{
\partial f}{\partial E}\,\,(\frac 1a\frac{da}{dt})(v^1v^1+v^2v^2+v^3v^3)
\label{TD3}
\end{equation}
or, since $v_\alpha v^\alpha =-c^2$,

\begin{equation}
\frac{Df}{Dt}=\frac E{m_oc^2}\frac{\partial f}{\partial
t}\,+m_o\frac{
\partial f}{\partial E}\,\,(\frac \theta 3)(\frac{E^2}{m_o^2c^2}-c^2)
\label{TD4}
\end{equation}
The factor  $\theta =\frac 3a\frac{da}{dt}$ is precisely a consequence of the geometrical features appearing in Eq. (\ref{TD2}), and it is usually identified with the divergence of the four-velocity. Since the spatial gradients vanish in a FRW space-time, emphasis should me made on the fact that the non-vanishing terms of this "divergence" come from the Christoffel symbols introduced in the covariant derivative, so that $\theta$ has a strictly geometrical origin. The solution of Eq (\ref{BGK1}), taking into account the
expression (\ref{TD4}) can be approximated up to second order in $\tau $
using the proposal:
\begin{equation}
f=f^{(0)}+\tau f^{(1)}+\tau ^2f^{(2)}+...  \label{sol1}
\end{equation}
Substituting Eq. (\ref{sol1}) into Eq. (\ref{TD4}) and
equating the coefficients of equal powers in constant $\tau $, up to second
order, we get:

\begin{equation}
f^{(1)}=-\left[ \frac E{m_oc^2}\frac{\partial f^{(0)}}{\partial
t}\,+m_o \frac{\partial f^{(0)}}{\partial E}(\frac \theta
3)(\frac{E^2}{m_o^2c^2} -c^2)\right]  \label{sol2}
\end{equation}
and
\begin{equation}
f^{(2)}=-\left[ \frac E{m_oc^2}\frac{\partial f^{(1)}}{\partial
t}\,+m_o \frac{\partial f^{(1)}}{\partial E}\,\,(\frac \theta
3)(\frac{E^2}{m_o^2c^2} -c^2)\right]  \label{sol3}
\end{equation}
Expressions (\ref{sol2}-\ref{sol3}) constitute our first and second order
corrections to the equilibrium distribution of the BGK equation in a flat
FRW space-time. For non-relativistic particles in a FRW metric, Eqs. (\ref
{sol2}-\ref{sol3}) reduce to:
\begin{equation}
f_{NR}^{(1)}=-\left[ \frac{\partial f^{(0)}}{\partial
t}\,+m_ou^2\frac{
\partial f^{(0)}}{\partial E}\,\,(\frac \theta 3)\right]  \label{sol4}
\end{equation}
\begin{equation}
f_{NR}^{(2)}=-\left[ \frac{\partial f^{(1)}}{\partial
t}\,+m_ou^2\frac{
\partial f^{(1)}}{\partial E}\,\,(\frac \theta 3)\right]  \label{sol5}
\end{equation}
The equilibrium function $f^{(0)}$ is, in this case, the Maxwell-Boltzmann
distribution:
\begin{equation}
f^{(0)}=n(\frac m{2\pi kT})^{\frac 32}e^{-\frac{mu^2}{2kT}}=n(\frac m{2\pi
kT})^{\frac 32}e^{-\frac E{kT}}  \label{sol6}
\end{equation}
where $n$ is the particle density, $m$ is the mass of one particle of the
simple system, $k$ is Boltzmann's constant and $T$ is the temperature.

If particles are conserved, then the continuity equation holds:
\begin{equation}
\frac{\partial n}{\partial t}+n\theta =0  \label{sol7}
\end{equation}
and the first term at the RHS of Eq. (\ref{sol4}) can be written as:
\begin{equation}
\frac{\partial f^{(0)}}{\partial t}=\frac{\partial
f^{(0)}}{\partial n}\frac{
\partial n}{\partial t}=-\theta f^{(0)}  \label{deriv1}
\end{equation}
In Eq. (\ref{deriv1}) $\theta$, usually identified with $\nabla \cdot \textbf{u}$ has, as emphasized before,  a purely geometrical origin; it is, indeed the surviving term of the covariant derivative of a four velocity $v_{;\alpha}^{\alpha}$. 
Straightforward calculations can now be carried, so that:
\begin{equation}
f_{NR}^{(1)}=(1+\frac{mu^2}{3kT})\theta f^{(0)}  \label{sol8}
\end{equation}
\begin{equation}
f_{NR}^{(2)}=\left[ -(1+\frac{5mu^2}{9kT}-\frac
2{27}\frac{m^2u^4}{k^2T^2} )\theta ^2+\theta \frac{d\theta
}{dt}\right] f^{(0)}  \label{sol9}
\end{equation}
These corrections obviously vanish for homogeneous, isotropic,
non-expanding space-times. The thermodynamical force inherent to
the deviations from the perfect distribution is, to first order,
the expansion scalar $\theta $. To second order (Burnett regime),
the generalized thermodynamical forces are, as expected, non-linear forces appearing in the linear constitutive relations. Here, due to the assumption on space-time, they turn out to be simply $ \theta ^2$ and
$\theta \frac{d\theta }{dt}$. The effect of these second order
expressions on the field equations will be the subject of the next
section.

\section{Stress-tensor and field equation}

The Einstein field equations with vanishing cosmological constant are given by:
\begin{equation}
G_\nu ^\mu =\kappa \,T_\nu ^\mu  \label{Einstein}
\end{equation}
where $G_\nu ^\mu $ is the Einstein tensor, $\kappa $ is the coupling
constant and $T_\nu ^\mu $ is the stress-energy tensor. The tensor $T_\nu
^\mu $ is related to the non-equilibrium distribution function, up to second
order, by:

\begin{equation}
T_\nu ^\mu \simeq m \int f^{(0)}v^\mu v_\nu \,dV+m \tau \int f^{(1)}v^\mu v_\nu
\,dV+m \tau ^2\int f^{(2)}v^\mu v_\nu \,dV  \label{stress2}
\end{equation}
Eq. (\ref{stress2}) is simply the ordinary stress-energy tensor written in terms of averages of the molecular velocities. The hydrodynamical velocity vanishes in the comoving frame. In this case, the first term of the R.H.S. of Eq. (\ref{stress2})  may be written as:
\begin{equation}
\stackrel{(0)}{T_\nu ^\mu }=\frac{4\pi }3m\,h_\nu ^\mu \int
f^{(0)}u^4\,du+\rho U^\mu U_\nu =\rho U^\mu U_\nu +P(\delta _\nu
^\mu +\frac{ U^\mu U_\nu }{c^2})=\rho U^\mu U_\nu +Ph_\nu ^\mu
\label{Stress2a}
\end{equation}
where $\rho $ is the mass-energy density, $P=nkT$ is the pressure, $h_\nu
^\mu =\delta _\nu ^\mu +\frac{U^\mu U_\nu }{c^2}$, and $U^\mu $ is the
hydrodynamic velocity given by:
\begin{equation}
U^\mu =\left[
\begin{array}{l}
0 \\
0 \\
0 \\
c
\end{array}
\right] ,\;U_\mu =\left[
\begin{array}{l}
0 \\
0 \\
0 \\
-c
\end{array}
\right]   \label{VFRW}
\end{equation}
The RHS of Eq. (\ref{Stress2a}) corresponds to the stress tensor of a
non-dissipative fluid. If we wish to include dissipation up to second order
in $\tau $ for non-relativistic particles,  we can use expressions (\ref
{sol8}-\ref{sol9}) . Using this information, and assuming that no
dissipation in the time axes occur, we can write for the first order
correction of the stress energy tensor:
\begin{equation}
\stackrel{(1)}{T_\nu ^\mu }=-\left[ 4\pi \tau m h_\nu ^\mu \int_o^\infty
\left( 1+\frac 13\frac{m_ou^2}{kT}\right) \,f^{(0)}\,u^4du\right] \theta
=-\eta _c\theta \,h_\nu ^\mu   \label{orden1}
\end{equation}
Following the rules of linear irreversible thermodynamics \cite{Groot}, the relationship between the induced current $ \stackrel{(1)}{T_\nu ^\mu } $ and the force $ \theta $ which again we emphasize arises from the inherent geometrical aspect of space-time, is $ \stackrel{(1)}{T_\nu ^\mu } = - \eta _{c} \theta h_\nu ^\mu $. $ \eta _{c} $, the transport coefficient is usually referred to in the literature as bulk viscosity. This is misleading, since $ \theta $ does not contain spatial gradients. For this reason, we shall refer to it as the "cosmological viscosity" $ \eta _{c} $, here explicitly given by:
\begin{equation}
\eta _{c}=4\pi \tau m  \int_o^\infty \left( 1+\frac
13\frac{m_ou^2 }{kT}\right) \,f^{(0)}\,u^4 du  \label{visco1}
\end{equation}
The second order contributions to the stress tensor can now be identified as:
\begin{eqnarray}
\stackrel{(2)}{T_\nu ^\mu } &=&-\left[ 4\pi \tau ^2 m 
h_\nu ^\mu \int_o^\infty \left( 1+\frac 59\frac{m_ou^2}{kT}+\frac
2{27}\frac{m_o^2u^4}{k^2T^2}\right)
\,f^{(0)}\,u^4 du\right] \theta ^2+  \label{ordenB} \\
&&\left[ 4\pi \tau ^2 m  h_\nu ^\mu \int_o^\infty \left( 1+\frac
13\frac{m_ou^2}{ kT}\right) \,f^{(0)}\,u^4du\right] \theta
\frac{d\theta }{dt} \nonumber
\end{eqnarray}
Thus, two Burnett transport coefficients arise, one for each
generalized thermodynamical force $\theta ^2$, $\theta
\frac{d\theta }{dt}$, namely:
\begin{equation}
B_1=\left[ 4\pi \tau ^2 m \int_o^\infty \left( 1+\frac
59\frac{m_ou^2 }{kT}+\frac 2{27}\frac{m_o^2u^4}{k^2T^2}\right)
\,f^{(0)}\,u^4 du\right] \label{B1}
\end{equation}
and
\begin{equation}
B_2=\left[ 4\pi \tau ^2 m \int_o^\infty \left( 1+\frac
13\frac{m_ou^2 }{kT}\right) \,f^{(0)}\,u^4 du\right]   \label{B2}
\end{equation}
and the field equations (\ref{Einstein}), in the Burnett regime, in flat FRW space-times, take the rather interesting  form
\begin{equation}
\frac 23\frac{d\theta }{dt}+\frac 5{27}\theta ^2=\kappa \,\left( p-\eta
_c\theta -B_1 \theta ^2+B_2 \theta \frac{d\theta }{dt}\right)   \label{FB1}
\end{equation}
\begin{equation}
\frac 13\theta ^2=\kappa \rho c^2  \label{FB2}
\end{equation}
These equations, clearly dependent on the mean free time $\tau $,
yield the dynamics of  simple fluid consisting of non-relativistic
particles in flat FRW metric in the Burnett regime. Future work will be
dedicated to the study of their solutions.

The order of magnitude of these coefficients is readily calculated. Substituting $ f^{(0)}$ one obtains that
\begin{equation}
\eta_{c}\approx 2 n k T \tau=\frac{B_1 }{\tau}  \label{Burn1}
\end{equation}
and
\begin{equation}
B_2\approx \frac{5}{3} n k T \tau^{2} \label{Burn2}
\end{equation}

The cosmological viscosity $\eta_{c}$ will be relevant when $\tau$ is of order one. Since for the Boltzmann dilute regime $\tau\sim 10^{-5} s- 10^{-8} s$, and for such values densities of the order $10^{-5}$, $10^{-19} g/cm^{3}$ should prevail, corresponding roughly to the threshold of the matter domination era in the universe.  Under these circumstances, the Burnett corrections which contain $\tau^{2}\sim10^{-16} s^{2}$ would require densities of the order of magnitude of $10^{-11} g/cm^{3}$, the density of matter right after nucleosynthesis, an era where radiation is still relevant  \cite{Turner}.  Further study of these consequences will be reported later.

\section{\protect\vspace{0.5cm}Discussion of the results}
In this paper we have used the BGK model  to describe the non-relativistic kinetic properties of a dilute gas in a FRW metric implying an isotropic and homogeneous universe. Two important results stem out of this analysis. In the first place the concept itself of transport coefficients is clarified. There existing no spatial gradients in the velocity, there will be no shear effects in the gas. Thus, ordinary viscosities, shear and volume, are not present. However, the contributions to the four dimensional covariant derivative of the velocity arising only from the geometric structure of space-time, embedded in the Christoffel symbols are present. They give rise to fluxes $ T_\nu ^\mu $ arising from the dynamical character of the fluid. In the case here studied, the force turns out to be the expansion scalar $ \theta $. As we have shown elsewhere \cite{Yo3}, there is a shear-like viscosity arising from geometry. Regretfully the nomenclature used in that paper is inappropriate, the coefficients then referred to are precisely $ \eta _{c} $ and its "shear-like" counterpart which should be called something else, \underline{not} shear viscosity.

The second important result is that as Eq. (\ref{orden1})-(\ref{FB2}) point out, there was  a Burnett-like hydrodynamic regime in the evolution of the universe which reflects itself in the field equations themselves, as well as in the transport properties of matter. Nevertheless, it did so at times when matter was still coupled to radiation, so that it may me useful to add to this analysis the behavior of radiation. This subject is, at present, one of study.


\begin{thebibliography}{1}
\bibitem[1]{Weinberg}  S. Weinberg; Ap.J. \textbf{168}, 175 (1971)
\bibitem[2]{GC} For a review see L.S. Garc\'{\i}a-Col\'{\i}n, F.J. Uribe and R.M. Velasco (to be published)
\bibitem[3]{Cow} S. Chapman and T.G. Cowling; "The mathematical theory of non-uniform gases", Cambridge Univ. Press, Cambridge, 3rd. Ed. (1970)
\bibitem[4]{Anderson1}J.L. Anderson in the Proc. of the Relativity Theory in the Midwest; M. Carmeli, S.I. Fickler and L. Witten, editors, Plenum Press N.Y. (1970) p 109
\bibitem[5]{Anderson2}J.L. Anderson and A.C. Payne Jr; Physica A (1976) \textbf{85} 261
\bibitem[6]{Berne} B.J. Berne and R. Pecora; Dynamic Light Scattering with Applications to Physics, Chemistry and Biology; Dover Publications, Minneola, N.Y. (2001)
\bibitem[7]{Mountain} R.Mountain, Rev. Mod. Phys. \textbf{38}, 205 (1996)
\bibitem[8]{Yip} S.Yip and J.P. Boon; Molecular hydrodynamics; Dover publications, Minneola, N.Y. (1980).
\bibitem[9]{yo}  A. Sandoval-Villalbazo; Physica A (2002) \textbf{313} 456
[cond-mat/0109274]
\bibitem[10]{Yo2} A. Sandoval-Villalbazo and L.S. Garc\'{i}a-Col\'{i}n (2002) [astro-ph/0207218]
\bibitem[11]{Gross} P.L. Bhatnagar, E.P. Gross and M. Krook; Phys. Rev.  (1954) \textbf{94} 511
\bibitem[12]{Spiegel}See also E.A. Spiegel and J.L. Thiffeault (2002)[astro-ph/0210185]
\bibitem[13]{Groot} S.R. de Groot and P. Mazur, "Non-equilibrium thermodynamics" (Dover publications, Minneola, N.Y. (1984)
\bibitem[14]{Yo3} A. Sandoval-Villalbazo and L.S. Garc\'{\i}a-Col\'{\i}n, Physica A (2000) \textbf{287} 307 
\bibitem[15]{Turner} E.W. Kolb and M.S. Turner, "The early universe", Addison-Weley Publ. Co. Reading Mass. (1990).
\end{thebibliography}
\end{document}